\documentclass[twoside]{mhd}
\usepackage{graphics,epsfig}

\title{Increasing the magnetic sensitivity of liquid crystals by rod-like magnetic nanoparticles}

\author{P. Kop\v{c}ansk\'y\inst{1}, N. Toma\v{s}ovi\v{c}ov\'a\inst{1}, T.T\'oth-Katona\inst{2}, N. \'Eber\inst{2},\\ 
M. Timko\inst{1}, V. Z\'avi\v{s}ov\'a\inst{1}, J. Majoro\v{s}ov\'a\inst{1}, M. Raj\v{n}ak\inst{1},\\ 
J. Jadzyn\inst{3},  X. Chaud\inst{4}}

\institute{$^{1}$Institute of Experimental Physics, Slovak Academy
of
Sciences, Watsonova 47, 04001 Ko\v{s}ice, Slovakia \\
$^2$Institute for Solid State Physics and Optics, \\ Wigner Research
Centre  for Physics,  Hungarian Academy of Sciences, \\ H-1525
Budapest, P.O.Box 49, Hungary \\
$^{3}$Institute of Molecular
Physics, Polish Academy of Sciences,
Smoluchowskiego 17,  60179 Poznan, Poland \\
$^{4}$High Magnetic Field Laboratory, CNRS, 25 Avenue des Martyrs,
\\ Grenoble,
 France}

\begin{document}
\maketitle

\begin{abstract}

Magnetic Fr\'eedericksz transition was studied in ferronematics
 based on the nematic liquid crystal
4-(trans-4'-n-hexylcyclohexyl)-isothiocyanatobenzene (6CHBT).
6CHBT was  doped with rod-like magnetic particles of different
size and volume concentration.  The volume concentrations of
magnetic particles in the prepared ferronematics were
$\phi_1$~=~10$^{-4}$ and $\phi_2$~=~10$^{-3}$. The structural
changes were observed by capacitance measurements that demonstrate
a significant influence of the concentration, the shape
anisotropy, and/or the size of the magnetic particles on the
magnetic response of these ferronematics.

\end{abstract}


\section{Introduction}


Liquid crystalline phases occur as additional, thermodynamically
stable states of matter between the liquid state and the crystal
state in some materials. They can be characterized by a long-range
orientational order of the molecules and, as a consequence, by an
anisotropy in their physical properties. Liquid crystals can be
oriented under electric or magnetic fields due to the anisotropy of
dielectric permittivity or diamagnetic susceptibility \cite{Gennes}.


One of the most important findings related to controlling liquid
crystals by external fields was the threshold behaviour in the
reorientational response of liquid crystals -- an effect described
by V.K. Fr\'eedericksz \cite{Free}, and named after him as
"Fr\'eedericksz transition". The dielectric permittivity anisotropy
of liquid crystals is  in general relatively large; thus driving
voltages of the order of a few volts are sufficient to control the
orientational response. Therefore, most of the liquid crystal
devices are driven by electric field. On the other hand, because of
the small value of the anisotropy of the diamagnetic susceptibility
($\chi_a$ $\sim$ 10$^{-7}$), the magnetic field H necessary to align
liquid crystals have to reach rather large values ($B$=$\mu_0$H
$\sim$ 1T), and therefore, liquid crystal applications using
magnetic fields are rather limited. Consequently, the increase of
the magnetic sensitivity of liquid crystals is an important
challenge, which can potentially broaden the area of applications
and may offer an opportunity to develop new devices.

Brochard and de Gennes \cite{BG} first suggested the idea that could
increase the magnetic sensitivity of liquid crystals theoretically.
According to them, colloidal systems called "ferronematics",
consisting of nematic liquid crystals doped with magnetic
nanoparticles in small concentrations, should respond to low
magnetic fields of the order of tens of Gauss. Such small
magnetic fields cannot affect the undoped nematics, however, they
may be sufficiently strong to align or rotate the magnetic moments
of the particles inside the ferronematic suspensions according to
the predictions. This realignment or rotation effect could then be
transferred to the host nematic through the coupling between the
nanoparticles and the liquid crystal molecules. One has to note here
that the realignment of the nematic host was assumed to be entirely
determined by the ferromagnetic properties of nanoparticles (not
affected by the intrinsic diamagnetic properties of the nematic),
since the theory \cite{BG} predicted a rigid anchoring with
{\textit{\textbf{n}} $\parallel$ {\textit{\textbf{m}}, where
{\textit{\textbf{n}} is the unit vector of the preferred direction
of the nematic molecules (director),  and {\textit{\textbf{m}} is
the unit vector of the magnetic the moment of the magnetic
particles.

The first experimental realization of ferronematic materials was
carried out by Chen and Amer \cite{Chen}. Later experiments on some
other ferronematics have indicated that besides the predicted
{\textit{\textbf{n}} $\parallel$ {\textit{\textbf{m}} condition, the
case of {\textit{\textbf{n}} $\perp$ {\textit{\textbf{m}} is also
possible. To bridge this gap between the theory and experiments,
Burylov and Raikher modified the theoretical description by
considering a finite value of the surface density of anchoring
energy $W$ at the nematic -- magnetic particle boundary
\cite{BR1,BR2,BR3,BR4}. The finite value of $W$, as well as the
parameter $\omega$= $Wd/2K$ ($d$ is the mean diameter of the
magnetic particles and $K$ is an orientational-elastic Frank
modulus), characterize the type of anchoring of nematic molecules on
magnetic particle's surface. For $\omega$ $>>$ 1 the anchoring is
rigid, while the soft anchoring is characterized by $\omega$ $\leq$
1 which (unlike the rigid anchoring) permits both types of boundary
conditions, \textit{\textbf{n}} $\parallel$ {\textit{\textbf{m}} and
{\textit{\textbf{n}} $\perp$ {\textit{\textbf{m}}.

So far the magnetic nanoparticles have established their wide range
of applications. The properties of magnetic nanoparticles
significantly depend on their size, shape and structure. Controlling
the shape and size of nanoparticles is one of the ultimate
challenges in modern material research. These magnetic particles can
be made so small that each particle becomes a single magnetic
domain, exhibiting abnormal magnetic properties, known as
superparamagnetism. Doping liquid crystals with low volume
concentration of nanoparticles has been shown to be a promising
method to modify the properties of liquid crystals. At such a low
amount, nanosized particles do not disturb significantly the liquid
crystalline order, hence producing a macroscopically homogeneous
structure. However, the particles can share  their properties with
the liquid crystal host, enhancing the existing properties, or
introducing some new properties for the composite mixtures.

\section{Experiment}

Two types of magnetic rod-like particles were prepared  through
hydrolysis of FeCl$_3$ and FeSO$_4$ solutions (molar ratio 2:1)
containing urea. To prepare the larger rod-like particles (sample A)
0.6756 g of FeCl$_3$~$\cdot$~6H$_2$O, 0.3426 g
FeSO$_4$~$\cdot$~7H$_2$O and 0.60 g (NH$_2$)$_2$CO were dissolved in
10~ml of purified, deoxygenated water. The product was added to a
flask with reflux condenser, which has been kept at 90-95$^\circ$C
for 12~h, and then cooled to room temperature {\cite{Xuebo}}. After
the synthesis particles were coated with oleic acid as a stabilizer.
In the synthesis of the smaller rod-like particles (sample B), at
first, the stabilizer (oleic acid) was ultrasonically dispersed in
water to form homogenous micelles. Then, FeCl$_3$~$\cdot$~6H$_2$O,
FeSO$_4$~$\cdot$~7H$_2$O were dissolved in the above solution. This
mixture was added to a flask with reflux condenser and has been
heated in water bath for 12 hours at 90-95~$^\circ$C during which a
dark precipitate has been formed. The sample has been cleaned
several times by purified and deoxygenated water, and then it has
been dried under low pressure at 50~$^\circ$C for 3~hours
{\cite{Lian}}. Figure~\ref{rods} shows transmission electron
microscopic images of the  prepared magnetic particles. The diameter
of the larger rod-like particles (sample A) was d$_A$=$(18 \pm
3)$~nm and their mean length L$_A$=$(400 \pm 52)$~nm. The mean
diameter and length of the smaller rod-like particles (sample B)
were d$_B$=$(10 \pm 1)$~nm and L$_B$=$(50 \pm 9)$~nm,
respectively. Consequently, the mean shape anisotropies of the
nanoparticles were approximately L$_A$/d$_A \approx 22$ and
L$_B$/d$_B \approx 5$ for the two types of samples.

\begin{figure}[h]
\begin{center}
{\bf (a)} \includegraphics[width=11.2pc]{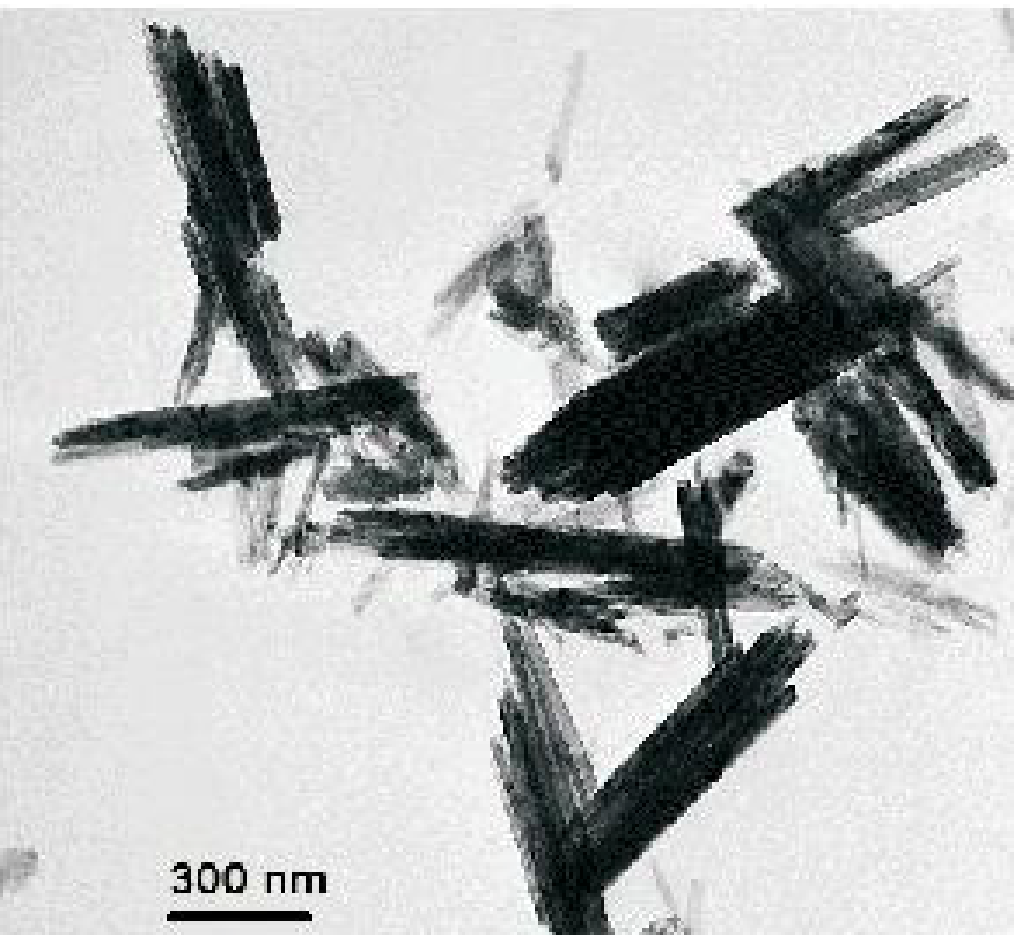} \hspace{1cm}
{\bf(b)} \includegraphics[width=12pc]{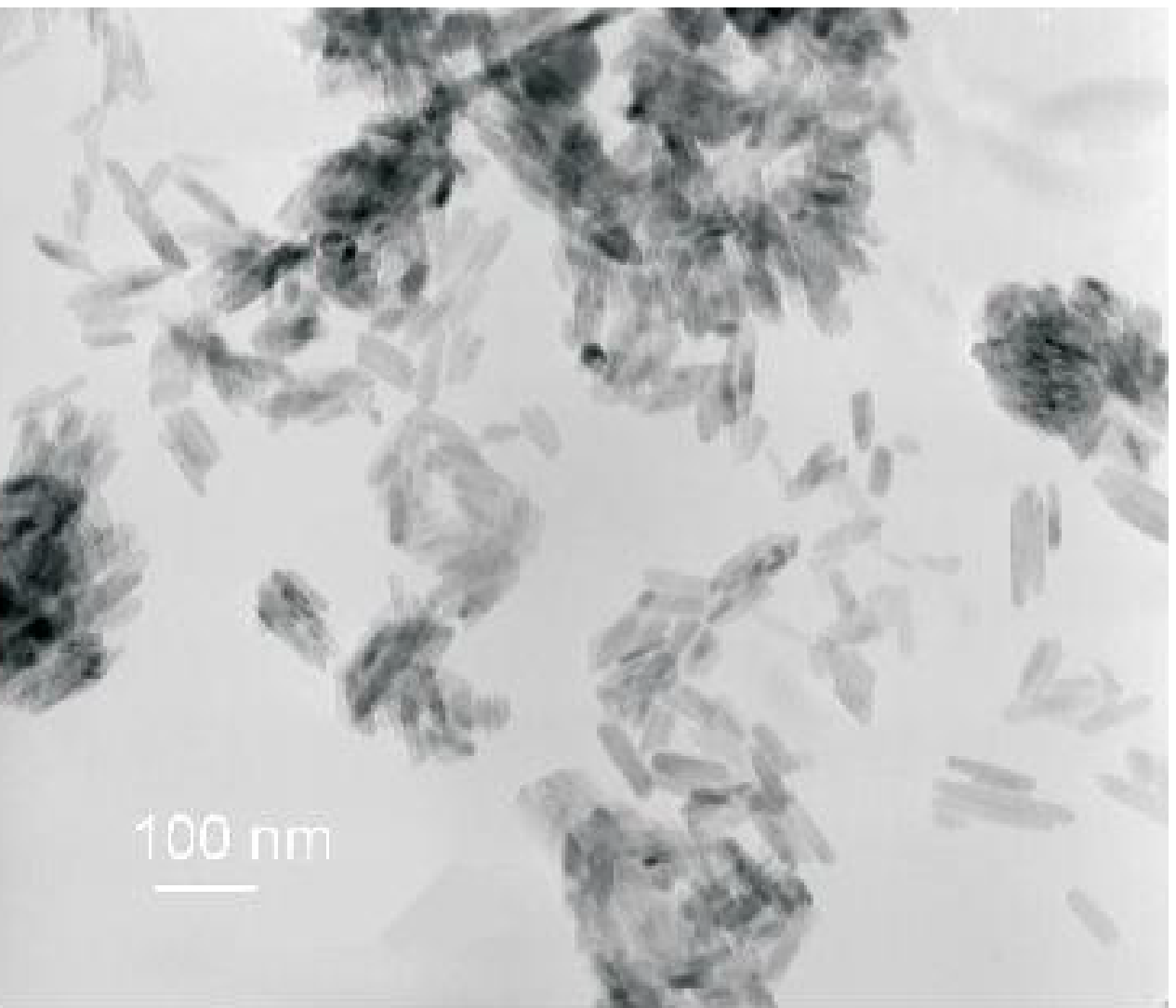} \caption{TEM image of
(a) the larger rod-like particles with mean diameter of 18~nm and
mean length of 400~nm; (b) the smaller rod-like particles with mean
diameter of 10~nm and mean length of 50~nm.} \label{rods}
\end{center}
\end{figure}

The ferronematic samples were based on the thermotropic nematic
4-(trans-4'-n-hexylcyclohexyl)-isothiocyanatobenzene (6CHBT), which
was synthetized and purified at the Institute of Chemistry, Military
Technical University, Warsaw, Poland. 6CHBT is a
low-temperature-melting enantiotropic liquid crystal with high
chemical stability {\cite{D}}. The phase transition temperature from
the isotropic liquid to the nematic phase has been found at
42.6~$^\circ$C. The doping  was  done by adding nanoparticles to the
liquid crystal in the isotropic phase under continuous stirring.

The structural transitions in the prepared samples were monitored by
capacitance measurements in a capacitor made of ITO-coated glass
electrodes. The capacitor with the electrode area of approximately
1~cm~$\times$~1~cm was placed into a regulated thermostat system,
the temperature of which was stabilized at 35$^\circ$C. The distance
between the electrodes (sample thickness) was $D$ = 5$\mu$m. The
capacitance was measured at the frequency of 1 kHz by the high
precision capacitance bridge Andeen Hagerling.

In the experiment the liquid crystal had a planar initial alignment,
i.e. the director was parallel to the capacitor electrodes, and the
magnetic field was applied perpendicular to them (see
Fig.~\ref{geo}). The dependence of the measured capacitance on the
external field reflects the reorientation of the nematic molecules.

\begin{figure}
\begin{center}
\includegraphics[width=10pc]{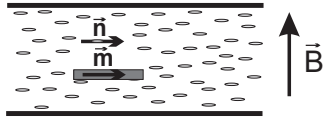}
\caption{Cross section of a planar cell with initially parallel
{\textit{\textbf{n}}} and {\textit{\textbf{m}}}.} \label{geo}
\end{center}
\end{figure}

\section{Results}

Results presented in work \cite{PK1} showed that doping with
magnetic particles shaped similarly to the  molecules of the host
liquid crystal is more effective and thus offers better perspectives
for ferronematics in applications where a magnetic field is
necessary to control the orientation of the liquid crystal. With the
aim to study  the influence of the size of the particles on the
magnetic response two kinds of rod-like magnetic particles (sample A
and sample B) were prepared as described above. Both ferronematics A
and B were based on the nematic 6CHBT, and were doped in two
different volume concentrations $\phi_1$~=~10$^{-4}$ and
$\phi_2$~=~10$^{-3}$.

Figure~\ref{size}(a) shows the magnetic Fr\'eedericksz transition in
pure 6CHBT and in ferronematics doped with larger (A) and smaller
(B) rod-like particles for both volume concentrations. It
demonstrates that the critical magnetic field $B_c$ of the
Fr\'eedericksz transition, i.e. the magnetic field that initiates
the reorientation of the director toward its direction, is shifted
to lower values with increasing the volume concentration, and that
$B_c$ is lower for larger particles than for smaller ones at a given
$\phi$. For all samples the critical magnetic field was determined
from the dependence of $(C-C_0)/(C_{max}-C_0)$ versus $B$, where
$C$, $C_0$ and $C_{max}$ are the capacitances at a given magnetic
field, at $B$ = 0, and at the maximum value of $B$, respectively.
$B_c$ was determined by linear extrapolation of data in
Fig.~\ref{size}(a). The obtained critical value of the magnetic
field for pure 6CHBT is 2.63~T. In ferronematics $B_c$ is lower, and
the values obtained for various samples are listed in Table 1. The
reduction of $B_c$ becomes larger if the concentration is increased
(in case of the same nanoparticle), as well as, if the nanoparticle
is larger (at the same concentration).

\begin{figure}[t]
\begin{center}
{\bf (a)} \includegraphics[width=16pc]{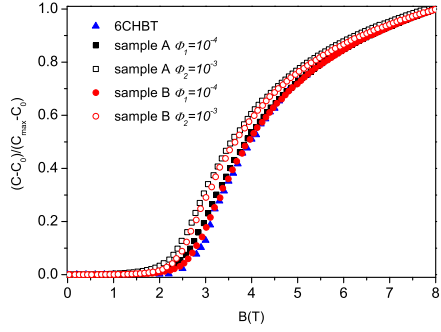} \\
{\bf (b)} \includegraphics[width=16pc]{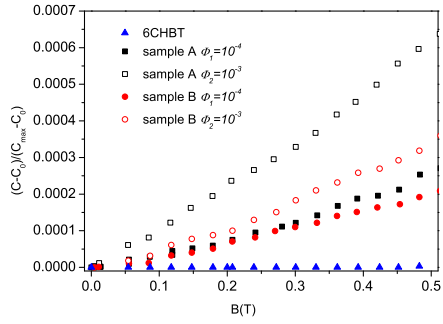} \caption{Reduced
capacitance versus magnetic field (a) for pure 6CHBT and for 6CHBT
doped with different rod-like particles and with different volume
concentrations of magnetic particles; (b) blowup of the low magnetic
field region for the same data.} \label{size}
\end{center}
\end{figure}

Observations of the structural transitions in ferronematics in
external field can be used for the determination of the type of
anchoring of nematic molecules on the surfaces of magnetic particles
as well as the surface density of the anchoring energy $W$ at the
nematic -- magnetic particle boundary. By means of the Burylov and
Raikher's expression for the free energy of the ferronematic
\cite{BR3} the formula for the critical magnetic field is

\begin{eqnarray}
B_{c}^2-B_{cFN}^2=\frac{2\mu_0W\phi}{\chi_ad} \label{eq:one},
\end{eqnarray}
where $B_c$  and  $B_{cFN}$ are the critical fields for the magnetic
Fr\'eedericksz transition of the pure liquid crystal and the
ferronematic, respectively, $d$ is the "characteristic size" of the
particles (the mean diameter), $\phi$ is the volume concentration of
magnetic particles in the liquid crystal, $\mu_0$ is the
permeability of vacuum and $\chi_a$ is the anisotropy of the
diamagnetic susceptibility of the liquid crystal (for 6CHBT $\chi_a$
= 4.805x10$^{-7}$ at 35$^\circ$C).

The calculated values of $W$ and the values of parameter $\omega$
are summarized in Table 1. $\omega$ has been calculated using the
same $K_1$~=~6.71~pN elastic constant for all ferronematics as for
the pure 6CHBT. In all cases $\omega$~$<$~1 that characterizes soft
anchoring of the nematic molecules on the surface of magnetic
particles.

\begin{table}[h]
\begin{center}
\begin{tabular}{cccc}

 sample & $B_{cFN}$ (T) & $W$ (Nm$^{-1}$) & $\omega$ \\
\hline
sample A $\phi_1$ & 2.39 & 4.14$\times$1$0^{-5}$  & 0.055 \\
sample A $\phi_2$ &2.12 & 8.34$\times$10$^{-6}$ & 0.011 \\
sample B $\phi_1$ & 2.52 & 1.08$\times$10$^{-5}$ & 0.008 \\
sample B $\phi_2$ & 2.25 & 3.54$\times$10$^{-6}$ & 0.003 \\
\hline
\end{tabular}
\caption{Critical magnetic fields measured in the ferronematics and
the calculated values of the surface density of the anchoring energy
$W$, and of the parameter $\omega$.}
\end{center}
\end{table}

In recent works by Podoliak et al. \cite{Podoliak}, and Buluy et al.
\cite{Buluy} both experimental and theoretical investigations have
been reported about the optical response of suspensions of
ferromagnetic nanoparticles in nematic liquid crystals on the
imposed magnetic field. The authors have measured a linear optical
response in ferronematics at very low magnetic fields (far below the
threshold of the Fr\'eedericksz transition).

A similar effect was also observed in our dielectric measurements in
samples doped with rod-like particles as it is demonstrated in
Fig.~\ref{size}(b). The figure provides a clear evidence for a
nearly linear magnetic field dependence of the capacitance in the
low magnetic field region.

\section{Conclusion}

We have demonstrated that both the threshold of the magnetic
Fr\'eedericksz transition and the dielectric response to low
magnetic fields (far below the Fr\'eedericksz transition) depend not
only on the volume concentration of the magnetic particles, but also
on the size of the particles. According to the results, the larger
is the particle, the bigger are the effects (larger decrease of the
threshold of the Fr\'eedericksz transition, and more pronounced
linear response to low magnetic fields). Since in our experiments
the larger particles have also a larger aspect ratio L/d, further
experiments are needed to clarify whether the volume size, the
linear size or the shape anisotropy (i.e. L/d) influences primarily
the magnitude of the effects.

The other challenging task is to explain the linear dielectric
response to low magnetic fields. To our present understanding,
within the framework of the Burylov and Raikher's continuum theory
\cite{BR1,BR2,BR3,BR4}, both the magnetic moment of magnetic
particle {\textit{\textbf{m}} and the presence of an initial
out-of-plane pretilt angle of the nematic director
{\textit{\textbf{n}} are necessary for a linear $C(B)$ dependence in
the low $B$ limit. A more detailed theoretical analysis is however,
needed to justify this assumption.

\section*{Acknowledgments}
This work was supported by the Slovak Academy of Sciences, in the
framework of CEX-NANOFLUID, projects VEGA 0045, the Slovak Research
and Development Agency under the contract No. APVV-0171-10, the
Ministry of Education Agency for Structural Funds of EU in the frame
of projects 26110230061, 26220120021 and 26220120033, the Grenoble High Magnetic
Field Laboratory (CRETA), and by the Hungarian Research Fund OTKA
K81250.





\end{document}